\documentclass[sigconf, screen, nonacm]{acmart}

\AtBeginDocument{%
  \providecommand\BibTeX{{%
    \normalfont B\kern-0.5em{\scshape i\kern-0.25em b}\kern-0.8em\TeX}}}

\usepackage{subcaption}
\usepackage{booktabs}
\usepackage{tabularx}

\usepackage{alphalph}
\makeatletter
\newcommand\footnoteref[1]{\protected@xdef\@thefnmark{\ref{#1}}\@footnotemark}
\makeatother

\usepackage{microtype}
\usepackage{listings}
\usepackage{tikz}
\usepackage{pifont}
\usetikzlibrary{shapes}

\lstdefinestyle{Cpp}{
    language=C++,
    keywordstyle=\color{blue},
    stringstyle=\color{red},
    commentstyle=\color{olive},
}

\lstdefinestyle{tinyCpp}{
    language=C++,
    basicstyle={\tiny\ttfamily},
    keywordstyle=\color{blue},
    stringstyle=\color{red},
    commentstyle=\color{olive},
}

\lstdefinestyle{MLIR}{
    sensitive=false,
    stringstyle=\color{red},
    string=[b]",
    commentstyle=\color{olive},
    comment=[l]{//},
    keywordstyle=[1]\color{blue},
    keywordstyle=[2]\color{violet},
    keywordstyle=[3]\color{brown},
    alsoletter={\%},
    keywords=[1]{sdfg, arith, memref, func, return},
    keywords=[2]{
        \%a, \%b, \%c, \%d, \%i, \%r,
        \@state_0, \@state_1, \@s0, \@s1, \@s2, \@s3, \@fName,
        \@init_2, \@constant_3, \@load_6, \@load_8, \@addi_9, \@return_11,
        \%arg0, \%arg1, \%arg2, \%arg3, \%arg4, \%arg5, \%arg6, \%arg7, \%arg8, \%arg9, \%arg10,
        \%c0, 
        \%0, \%1, \%2, \%3, \%4, \%5, \%6, \%7, \%8, \%9, \%10,
        \%alloc, \%alloca, \%c0_i32, \%c0_i8
    },
    keywords=[3]{i32, index, i8, i64, i1},
}

\lstdefinestyle{tinyMLIR}{
    sensitive=false,
    basicstyle={\tiny\ttfamily},
    stringstyle=\color{red},
    string=[b]",
    commentstyle=\color{olive},
    comment=[l]{//},
    keywordstyle=[1]\color{blue},
    keywordstyle=[2]\color{violet},
    keywordstyle=[3]\color{brown},
    alsoletter={\%},
    escapechar=\$,
    keywords=[1]{sdfg, arith, memref, func, return},
    keywords=[2]{
        \%a, \%b, \%c, \%d, \%i, \%r,
        \@state_0, \@state_1, \@s0, \@s1, \@s2, \@s3, \@fName,
        \@init_2, \@constant_3, \@load_6, \@load_8, \@addi_9, \@return_11,
        \%arg0, \%arg1, \%arg2, \%arg3, \%arg4, \%arg5, \%arg6, \%arg7, \%arg8, \%arg9, \%arg10,
        \%c0, 
        \%0, \%1, \%2, \%3, \%4, \%5, \%6, \%7, \%8, \%9, \%10,
        \%alloc, \%alloca, \%c0_i32, \%c0_i8
    },
    keywords=[3]{i32, index, i8, i64, i1},
}

\lstdefinestyle{tinyASM}{
    sensitive=false,
    basicstyle={\tiny\ttfamily},
    stringstyle=\color{red},
    string=[b]",
    commentstyle=\color{olive},
    comment=[l]{//},
    keywordstyle=[1]\color{blue},
    keywordstyle=[2]\color{violet},
    keywordstyle=[3]\color{brown},
    alsoletter={\%},
    keywords=[1]{main},
    keywords=[2]{
        \%rax, \%eax, \%edi, \%rcx, \%rdi, @PLT
    },
    keywords=[3]{2, 80000000000},
}

\newcommand{\macsection}[1]{\textbf{\textit{#1.}}~~}

\begin{document}

\title{MLIR-Smith: A Novel Random Program Generator for Evaluating Compiler Pipelines}

\author{Berke Ates}
\email{beates@student.ethz.ch}
\affiliation{%
  \institution{ETH Zurich}
  \country{Switzerland}
}

\author{Filip Dobrosavljević}
\email{dofilip@student.ethz.ch}
\affiliation{%
  \institution{ETH Zurich}
  \country{Switzerland}
}

\author{Theodoros Theodoridis}
\email{theodoros.theodoridis@inf.ethz.ch}
\affiliation{%
  \institution{ETH Zurich}
  \country{Switzerland}
}

\author{Zhendong Su}
\email{zhendong.su@inf.ethz.ch}
\affiliation{%
  \institution{ETH Zurich}
  \country{Switzerland}
}


%
\begin{abstract}
    Compilers are essential for the performance and correct execution of software and hold universal relevance across various scientific disciplines. 
    Despite this, there is a notable lack of tools for testing and evaluating them, especially within the adaptable Multi-Level Intermediate Representation (MLIR) context.
    This paper addresses the need for a tool that can accommodate MLIR's extensibility, a feature not provided by previous methods such as Csmith.
    Here we introduce MLIR-Smith, a novel random program generator specifically designed to test and evaluate MLIR-based compiler optimizations.
    We demonstrate the utility of MLIR-Smith by conducting differential testing on MLIR, LLVM, DaCe, and DCIR, which led to the discovery of multiple bugs in these compiler pipelines.
    The introduction of MLIR-Smith not only fills a void in the realm of compiler testing but also emphasizes the importance of comprehensive testing within these systems.
    By providing a tool that can generate random MLIR programs, this paper enhances our ability to evaluate and improve compilers and paves the way for future tools, potentially shaping the wider landscape of software testing and quality assurance.
\end{abstract}

\maketitle
\section{Introduction}
The performance of programs is significantly enhanced by compiler optimizations, which employ a variety of techniques such as dead code elimination, loop unrolling, and common sub-expression elimination~\cite{muchnick1997advanced}.
The effectiveness of these optimizations and the correctness of compilers can be evaluated through fuzz and differential testing using randomly generated programs. This approach was demonstrated by Yang et al.~\cite{Csmith} using Csmith, a tool specifically designed for generating random C programs.

The Multi-Level Intermediate Representation (MLIR)~\cite{mlir} project was initiated with the objective of mitigating the intricacies involved in compiler construction. MLIR serves as a unified infrastructure designed to alleviate software fragmentation, enhance compilation performance on heterogeneous hardware, and drastically cut down the expenditure involved in the development of domain-specific compilers. Moreover, it provides assistance in establishing connections between pre-existing compilers.

Polygeist~\cite{polygeistPACT} possesses the ability to produce MLIR code from C code. However, given its novelty, its robustness is not yet fully reliable for randomly generated C code. Additionally, MLIR is capable representing operations that surpass the instruction set of the C programming language. This presents a challenge as there is currently no dedicated tool for generating random MLIR programs, which is essential for comprehensive testing and evaluation of MLIR-based compilers and  optimizations.

\begin{figure}[t]
    \centering
    \vspace{0.8em}
    \includegraphics[width=.9\linewidth]{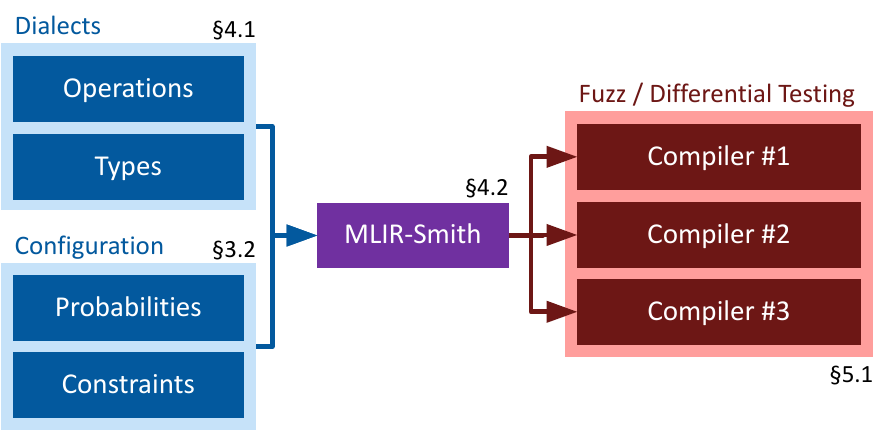}
    \vspace{-1em}
    \caption{Overview of MLIR-Smith.}
    \vspace{-1.5em}
    \label{fig:introduction_overview}
\end{figure}

To overcome this challenge, we introduce MLIR-Smith, a novel random program generator for MLIR. MLIR-Smith facilitates the generation of random MLIR programs, enabling a thorough evaluation of the optimization capabilities of MLIR-based compilers and ensuring the correctness and effectiveness of the compiled code.

We demonstrate the utility of MLIR-Smith by implementing a test harness that uses dead code elimination (DCE) techniques to evaluate the optimization capabilities of MLIR~\cite{mlir}, LLVM~\cite{llvm}, DaCe~\cite{dace}, and DCIR~\cite{mlir-dace}. DCE is a well-established method for analyzing the optimization efficiency of compilers and identifying missed opportunities for code optimization~\cite{10.1145/3503222.3507764}. In addition, we use MLIR-Smith to uncover bugs in MLIR, LLVM, DaCe, and DCIR through differential testing.

The paper makes the following contributions:
\begin{itemize}
    \item We design and implement MLIR-Smith, a random program generator that utilizes a minimal interface with operations, specifically for MLIR.
    \item We address the current limitations in MLIR testing infrastructure that inhibit effective compiler testing.
    \item We demonstrate the effectiveness of MLIR-Smith through a series of evaluations, showcasing its potential to enhance the field of compiler research and testing.
\end{itemize}

\section{Background}
We begin by discussing the foundational approaches that our work is built upon, specifically the MLIR~\cite{mlir} and DaCe~\cite{dace} frameworks, the DCIR pipeline~\cite{mlir-dace}, and Csmith~\cite{Csmith}.

\subsection{MLIR}
LLVM~\cite{llvm} and JVM~\cite{jvm} are examples of traditional compiler systems that rely on a single level of abstraction. These frameworks translate programs from one language into an intermediate representation (IR), where all necessary transformations, improvements, and verifications are made. The goal of MLIR~\cite{mlir} is to make the creation and interoperability of many IRs simpler by providing a framework that enables numerous levels of abstraction to coexist.

An intermediate representation in MLIR is called a \emph{dialect}. Specialized operations and data types are available in dialects, along with conversion points for lowering from higher-level IRs and converting between IRs of equal level. Although MLIR comes with a number of standard dialects, users are free to create their own unique dialects. The majority of dialects provide lowering passes one level down, reducing redundant code in compiler infrastructure development.

The expanding use of MLIR by front end and back end projects highlights the significance of this technology for compilers. For instance, the front end project Flang~\cite{flang} converts FORTRAN code into MLIR dialects. On the back end, a recent performance assessment by Katel et al.~\cite{mlir-gpu} compared MLIR to CUBLAS~\cite{cublas} to demonstrate that MLIR can produce GPU code that is competitive with well-known libraries for matrix-matrix multiplications.

Due to MLIR's rising popularity, there is a high demand for reliable testing tools, which are crucial for ensuring the correct implementation of dialects.

\subsection{DaCe}
A multitude of optimization techniques are designed to improve data movement, yet many are specifically tailored to distinct hardware systems or individual hardware configurations. Concurrently, the requirement for high-level programs to account for data movement often leads to an increase in code complexity.

By providing a Stateful DataFlow multiGraph (SDFG)~\cite{dace} representation, the DaCe programming framework provides a solution to this issue. The main goal of this IR is to comprehend and enhance data flow. DaCe offers front ends that convert Python, Octave, or C programming code into the SDFG IR. Additionally, it provides a transformation API on the IR that efficiently divides the responsibilities of the performance engineer and developer.
Applications in a variety of fields have benefited from the use of DaCe, including weather and climate models~\cite{stencilflow,fv3}, sparse linear algebra in quantum transport simulation~\cite{omen}, graph analytics~\cite{dace}, and comprehensive neural network optimization in deep learning~\cite{daceml}.

The DaCe framework's optimization method starts with an SDFG simplification pass that widens pure data-flow areas and eliminates superfluous memory allocation, much like the compiler's \texttt{-O1} flag. After that, automatic heuristic optimizations (\texttt{-O2}) and sometimes manual transformation application by performance engineers are used.
The SDFG IR is a control-flow graph, represented by a state machine, consisting of data-flow acyclic multigraphs. SDFGs distinguish between the usage of data in computational nodes (referred to as \textit{tasklets}) and the use of data containers and data movement (expressed as data-flow edges). As a result, the IR may describe data mobility directly, and it can even use data dependencies to decide the execution order. Explicit control-flow is only employed in cases, where inferring data-flow is not possible.

However, passes that are often implemented by control-centric IRs, such as loop-invariant code motion or general common subexpression elimination, cannot be represented by SDFGs by nature. The tasklet is regarded as an atomic unit, hence it's not possible to check its contents for changes.
Additionally, several design patterns, such as parametric-depth recursion or dynamic pointers, which might be well handled by other IRs, cannot be succinctly stated in SDFGs.

In order to enable optimizations on both ends, Ben-Nun et al.~\cite{mlir-dace} built a bridge between SDFG and MLIR.

\subsection{DCIR}
In a recent development, Ben-Nun et al.~\cite{mlir-dace} introduced a novel compiler pipeline, the Data-Centric Intermediate Representation (DCIR). DCIR uniquely combines the optimization capabilities of MLIR~\cite{mlir} and DaCe~\cite{dace}, to tap into previously unexplored optimization possibilities. DCIR accomplishes this by introducing a novel dialect, the SDFG dialect, which closely mirrors SDFGs. This is complemented by conversion and translation passes that bridge the gap between MLIR and DaCe.

The promising results demonstrated by DCIR, particularly in enhancing dead code elimination (DCE) capabilities in specific scenarios, hint at the potential for substantial benefits in a broader range of cases. Notably, the authors presented an example involving large arrays and nested loops, which DCIR was able to eliminate entirely, a feat not achieved by the compilers it was compared against, namely GCC~\cite{gcc}, Clang~\cite{clang}, DaCe, and MLIR.


\subsection{Csmith}
Traditional testing methods for compilers often fall short in terms of coverage and diversity of test cases. Csmith~\cite{Csmith}, a random program generator for C, addresses this issue by generating random, well-formed C programs. This allows for more comprehensive testing of compiler optimizations and transformations, as well as the identification of compiler bugs that might be missed by conventional testing methods.
Csmith generates programs by randomly selecting from a set of predefined operations and constructs, ensuring that the generated programs are syntactically correct and free of undefined behaviors. This is achieved through a combination of random decisions and safety checks, which ensure that the generated programs do not contain undefined behaviors that could lead to non-deterministic execution.
The utility of Csmith as a tool for uncovering compiler bugs and assessing the performance of compiler optimizations has been well-documented in previous research~\cite{Csmith}. Its ability to generate a diverse set of test cases makes it a valuable tool for compiler testing. However, Csmith is limited to generating C programs, which restricts its applicability to compilers and frameworks that support C.

While it is possible to use Csmith in conjunction with a tool like Polygeist~\cite{polygeistPACT} to generate MLIR~\cite{mlir} code, this approach has its limitations. Polygeist has demonstrated its proficiency in translating scientific applications, such as those found in the Polybench suite~\cite{polybench}. However, when tasked with translating C programs generated by Csmith, it becomes evident that Polygeist's stability is insufficient for this purpose. Furthermore, this approach would limit the scope of the generated MLIR programs to that of C code.

To address these limitations and to enable more comprehensive testing of MLIR, we propose the development of MLIR-Smith, a tool inspired by Csmith but specifically designed for MLIR.

\section{Approach}\label{sec:approach}
This section explains the intricate design choices behind MLIR-Smith and additionally gives an overview of the structure and design trade-offs of MLIR-Smith.

\subsection{Design Goals}
MLIR is designed to be a flexible and extensible compiler infrastructure, allowing the incorporation of new dialects and transformations. An intricately designed tool for the MLIR framework embraces this extensibility. Consequently, the main design goal of MLIR-Smith is to be a highly extensible tool for MLIR in which developers can easily integrate new operations from new dialects.

An additional design goal of MLIR-Smith is to accommodate the specific requirements of users, recognizing that they may have unique needs or prefer to focus on testing a subset of operations. By providing a high degree of configurability, MLIR-Smith empowers users to customize crucial aspects, such as the probabilities assigned to operations or the depth of nested blocks. This flexibility enables users to tailor the system to meet their distinct needs and preferences.

\subsection{Structure}\label{sec:structure}
MLIR-Smith is structured to seamlessly integrate within the MLIR environment, aligning with established MLIR tools and conventions. Following the structural paradigm of other MLIR tools like \texttt{mlir-opt} and \texttt{mlir-reduce}, MLIR-Smith maintains familiarity in its construction, facilitating its usage and development alongside other MLIR utilities. MLIR-Smith effectively utilizes the core capabilities of MLIR, while also introducing new utility functions to enhance usability.

MLIR-Smith is comprised of two distinct components, each serving a crucial role in its functionality. The core of MLIR-Smith initializes the configuration and sets up the initial program structure, consisting of the MLIR module and its \texttt{main()} function. Alongside this core component, MLIR-Smith incorporates a complementary utility class that extends the features of the internal MLIR operation builder and augments the overall functionality of MLIR-Smith. For example, the utility class provides helper functions that help generating blocks with requested return types. Moreover, developers gain access to helper functions to sample and generate operands of specific types.

Contrary to Csmith where the behavior and requirements of the generated C operations are known, MLIR-Smith has to adaptively select and insert possibly new and undocumented operations from new dialects. MLIR-Smith is oblivious to the semantics, constraints and requirements of these operations. Therefore, MLIR-Smith shifts the responsibility of generating operations to the respective developer of the dialect by delegating the generation process to the respective operations. Through this approach, we not only ensure compliance with the specific constraints of the generated operations but also embrace and leverage the inherent extensibility of the framework. To allow the registration of new operations, MLIR-Smith employs a robust mechanism by leveraging the MLIR dialect registry and identifying operations with a newly defined trait. While the dialect registry in MLIR is a centralized repository that stores information about the registered dialects and their associated operations, traits serve as annotations that provide additional information about the characteristics of operations. 

Alongside the trait, MLIR-Smith implements an interface known as \texttt{GeneratableOpInterface}. This interface specifies function declarations for generatable operations, thereby establishing a standardized structure. This interface is accompanied by a trait that simplifies the addition of operations to the generatable set, while also providing enhanced search capabilities. 
To enable seamless integration of newly developed operations into the program workflow, MLIR-Smith utilizes the dialect registry to search for operations with this specific trait. By leveraging the trait-based approach, MLIR-Smith ensures that only operations suitable for program generation are considered, facilitating a targeted and efficient generation process.

\subsection{Design Trade-offs}\label{sec:guarantees}
A number of design trade-offs have been carefully considered to strike a balance between flexibility and efficiency in MLIR-Smith. The specifics of these design trade-offs are detailed in the following paragraphs.

\macsection{No Guarantee of Termination}
Even though it is feasible to generate random programs that always terminate, Yang et al.~\cite{Csmith} suggest that this would limit the expressiveness of the generator too much. Moreover, Yang et al.~\cite{Csmith} argue that always-terminating tests cannot find compiler bugs that wrongfully terminate a non-terminating program. In other words, if the tests are limited to scenarios where programs always terminate, any bugs that cause non-terminating programs to erroneously terminate will remain undetected. Consequently, we have opted to adopt this design philosophy, offering no guarantee of termination. To manage potential non-terminating programs, we have implemented the use of timeouts.

\macsection{No Ground Truth}
Similar to Csmith~\cite{Csmith}, MLIR-Smith operates in a domain where the notion of a "ground truth" is not applicable. In the context of program generation and compiler testing, a ground truth represents a definitive correctness criteria against which the generated programs or compiler output can be compared to. However, due to the random and diverse nature of the programs generated by MLIR-Smith, establishing a universal ground truth for each generated program is impractical. Instead, MLIR-Smith is designed to be used in a differential test setting where the outputs of different compilers are compared. Despite this limitation, MLIR-Smith retains its usability for testing crashes in single compilers or optimization passes.

\macsection{Type safety}
In order to uphold array safety, MLIR-Smith disallows the generation of dynamic sizes and enforces static dimensions at compile-time. Specifically, MLIR-Smith generates \texttt{memref} types without strided access patterns nor affine maps and with a maximum of three dimensions, while employing static dimension sizes that can reach up to $100000$. This deliberate design decision enables MLIR-Smith to generate indices within the allowable range for each dimension, thereby reducing the occurrence of runtime errors related to array bounds. 
\section{Implementation}
In the following, we present a detailed description of the MLIR-Smith implementation including the process of selecting and building operations. 

\subsection{Supported Operations}
Since our work focuses on the evaluation of the four compiler frameworks MLIR, LLVM, DaCe, and DCIR, we specifically implement the subset of MLIR dialects supported by all frameworks. This enables us to generate DCIR-supported MLIR code and allows us to apply differential testing to all four compiler frameworks. A summary of the implemented operations can be found in Table~\ref{table:approach_implops}. As we can see, a subset of structured control flow (\emph{scf}) operations, such as \texttt{scf.for}, \texttt{scf.if}, and \texttt{scf.while}, as well as all instructions of the arithmetic (\emph{arith}) and math (\emph{math}) dialects are implemented. While the \emph{arith} dialect implements basic integer and floating point operations, the \emph{math} dialect contains operations beyond simple arithmetics. Lastly, DCIR supports a subset of the memory reference (\emph{memref}) dialect, which provides operations for the creation and manipulation of multi-dimensional memory references. The generated \emph{memref} types are subject to the constraints mentioned in \S\ref{sec:guarantees}.

\begin{table}[t]
  \centering
  \footnotesize
  \caption{A summary of the dialects and its operations that are generated by MLIR-Smith.}
  \vspace{-1em}
    \begin{tabularx}{\linewidth}{@{} p{3em} *1{>{\raggedright\arraybackslash}X} @{}}
        \toprule
            \textbf{Dialect} & \textbf{Implemented Operations} \\
        \midrule
            \texttt{func} & \texttt{func.return}, \texttt{func.call} \\
            
            \addlinespace[0.5em]
            \texttt{scf} & \texttt{scf.if}, \texttt{scf.for},\texttt{scf.while}, \texttt{scf.condition} and \texttt{scf.yield} \\
            
            \addlinespace[0.5em]
            \texttt{arith} & All operations implemented. \\
            
            \addlinespace[0.5em]
            \texttt{math} & All operations implemented. \\
            
            \addlinespace[0.5em]
            \texttt{memref} & \texttt{memref.alloc}, \texttt{memref.alloca}, \texttt{memref.load}, \texttt{memref.store}, \texttt{memref.dealloc} and \texttt{memref.copy} \\
        \bottomrule
    \end{tabularx}
  \label{table:approach_implops}
  \vspace{-1em}
\end{table}

\subsection{Program Generation}
Since Csmith has proven its effectiveness in generating diverse and challenging test cases, MLIR-Smith closely follows the core principles used to generate C programs in Csmith. Particularly, the random program generation process follows a top-down approach, starting at the program's \texttt{main()} function~\cite{Csmith}. Fig.~\ref{fig:approach_mlirsmith} shows a simplified overview of the \texttt{mlirSmithMain()} function, implementing the core of the program generation process. 

\begin{figure}[t]
    \centering
\begin{lstlisting}[style=tinyCpp]
LogicalResult mlir::mlirSmithMain(int argc, char **argv,
                                  DialectRegistry &registry) {
  // Parse command-line options.
  [...]

  // Load configuration.
  [...]

  // Create OpBuilder.
  GeneratorOpBuilder builder(&ctx, config);

  // Insert top-level module.
  ModuleOp module = ModuleOp::build(builder);
  builder.setInsertionPointToStart(module.getBody());

  // Insert main function.
  FunctionType funcType = builder.getFunctionType({}, {I32});
  func::FuncOp func = func::FuncOp::build(builder, "main", funcType);

  // Generate main function body.
  builder.setInsertionPointToStart(func.addEntryBlock());
  builder.generateBlock(/*ensureTerminator=*/true, /*requiredTypes=*/{I32});

  // Output the result.
  module.print();
  return success();
}
\end{lstlisting}
    \vspace{-0.5em}
    \caption{Simplified overview of the code generation process in MLIR-Smith.}
    \vspace{-1em}
    \label{fig:approach_mlirsmith}
\end{figure}

In the first step, MLIR-Smith's command line options are parsed. Users can specify the configuration file and seed to use, while additionally choosing an output directory for the generated code. MLIR-Smith further provides an option to dump the current configuration. In the next step, the configuration for the program generation is loaded. If no configuration file was specified, MLIR-Smith uses its default configuration values, which are omitted for brevity here. Further, an instance of the utility class \texttt{GeneratorOpBuilder} is created. \texttt{mlirSmithMain()} uses this instance to set up the initial program structure by building the MLIR module and the program's \texttt{main()} function. A block is hereinafter generated by \texttt{GeneratorOpBuilder} and inserted to the beginning of the \texttt{main()} function. In the following we describe the intricate details of the \texttt{GeneratorOpBuilder} class.

\macsection{Building Mechanism}
To enhance the control over the operation building process, we introduce the \texttt{GeneratorOpBuilder} class. This class extends the \texttt{OpBuilder} utility from the MLIR ecosystem, which is designed to streamline the construction of operations. The \texttt{OpBuilder} effectively manages the current block and region context during construction, maintaining the current insertion point within the MLIR function or block. This makes it straightforward to add operations at the correct locations.

The \texttt{GeneratorOpBuilder} expands upon this functionality by offering additional utility functions. These include the ability to sample operands and populate empty blocks. One key function provided by \texttt{GeneratorOpBuilder} is \texttt{generateBlock()}, which randomly generates operations within a block. If required, \texttt{generateBlock()} ensures the presence of a terminator and the provision of the necessary types for this terminator.

By extending the \texttt{OpBuilder} class, \texttt{GeneratorOpBuilder} can override certain functions, such as the \texttt{create()} function, to enforce specific constraints. These constraints can include limiting the depth of nested regions or the number of operations in a single block. Additionally, the \texttt{GeneratorOpBuilder} ensures the correct placement of the insertion point and implements a limited rollback feature. This feature allows it to reverse changes made by unsuccessful operation generations, thereby enhancing the control over the building process. This design also provides a flexible foundation for future extensions to the builder, such as the addition of further constraints or utilities.

\begin{figure}[t]
    \centering
\begin{lstlisting}[style=tinyCpp]
LogicalResult arith::AddIOp::generate(GeneratorOpBuilder &builder) {
  llvm::SmallVector<Type> possibleTypes = getGeneratableTypes(builder);

  while (!possibleTypes.empty()) {
    unsigned idx = builder.sampleUniform(possibleTypes.size() - 1);
    Type resultType = possibleTypes[idx];

    llvm::Optional<Value> lhs = builder.sampleValueOfType(resultType);
    llvm::Optional<Value> rhs = builder.sampleValueOfType(resultType);

    if (!lhs.has_value() || !rhs.has_value()) {
      Type *it = llvm::find(possibleTypes, resultType);
      if (it != possibleTypes.end())
        possibleTypes.erase(it);
      continue;
    }

    OperationState state(builder.getUnknownLoc(),
                         arith::AddIOp::getOperationName());
    arith::AddIOp::build(builder, state, lhs.value(), rhs.value());
    if (builder.create(state) != nullptr)
      return success();

    Type *it = llvm::find(possibleTypes, resultType);
    if (it != possibleTypes.end())
      possibleTypes.erase(it);
  }

  return failure();
}
\end{lstlisting}
    \vspace{-0.5em}
    \caption{Overview of the \texttt{arith.addi} generation function.}
    \vspace{-1em}
    \label{fig:approach_genop}
\end{figure}

\macsection{Operation Generation}
For an operation to be recognized as generatable by MLIR-Smith, the \texttt{GeneratableOpInterface} trait must be appended to the operation's trait list. Moving forward, each operation must define two separate functions, namely \texttt{generate()} and \texttt{getGeneratableTypes()}. While \texttt{generate()} handles the process of building the respective operation, \texttt{getGeneratableTypes()} returns a list of types that can be generated by the operation's generation function. \texttt{getGeneratableTypes()} is utilized in a number of helper functions implemented in \texttt{GeneratorOpBuilder}. For example, \texttt{generateValueOfType()} uses the \texttt{getGeneratableTypes()} function to search for operations that generate the requested return type.
Another use case is found within the implementation of \texttt{sampleTypes()}. Its task is to sample from a geometric distribution of available types in the current position of the block. It essentially utilizes \texttt{getGeneratableTypes()} to collect all unique generatable types from the available operations. Nevertheless, the utilization of \texttt{getGeneratableTypes()} extends beyond its role in \texttt{GeneratorOpBuilder}, as it may also be employed within the \texttt{generate()} function. 

Fig.~\ref{fig:approach_genop} illustrates the code of generating the \texttt{arith.addi} operation. As depicted, much of the functionality is encapsulated within the \texttt{GeneratorOpBuilder}. This abstraction simplifies the process for developers, making it straightforward to integrate new dialects into MLIR-Smith for testing.
While MLIR accommodates integers of any size, we have chosen to support only integers of width $1, 8, 16, 32,$ and $64$ bits. This decision is based on the observation that these are the most commonly used integer widths. Similarly, for floating point types, we have opted to support widths of $16, 32,$ and $64$ bits.

In every loop iteration of the \texttt{arith.addi} generation function, a result type is sampled from the list of generatable types and two operands of the same type are then sampled using the utility function \texttt{sampleValueOfType()} of \texttt{GeneratorOpBuilder}. It is important to note that some operations require operands of specific values and types and it is the developer's responsibility to ensure that the correct operands are sampled according to their distinct preferences.

In the case of \texttt{arith.addi}, the type of the operands must coincide with the result type and no special requirements are enforced on the operands. If the operation building process succeeds, the operation is inserted at the current location. However, if no operands of the specified type could be found or the creation of the operation failed due to violating limitations, the building process fails. In this case, we remove the result type from the set of possible generatable types and try again in a new loop iteration. If all loop iterations fail in building an operation, a failure signal is returned. \texttt{GeneratorOpBuilder} handles a failure by advancing to the next available operation.



\section{Evaluation}
In this section, we showcase the practical effectiveness of MLIR-Smith as a tool for uncovering bugs and missed optimizations in the MLIR, LLVM, DaCe, and DCIR compiler pipelines, thereby demonstrating its value in enhancing the reliability and capability of these systems.

\subsection{Experimental Setup}\label{sec:experimental_setup}
Our evaluation leverages MLIR-Smith with the configuration detailed in Table~\ref{table:evaluation_configparams} to generate MLIR programs.
To enhance the effectiveness of our bug discovery process, we have chosen to limit nested regions to a depth of $4$ and the number of operations in a block to $50$, based on the premise that bugs are often revealed through relatively small examples. Further, we have assigned high probabilities to \texttt{scf.if}, \texttt{scf.for}, and \texttt{scf.while} operations, as these operations are expected to be more prone to unveiling bugs and missed optimizations. Finally, we have deactivated a subset of \texttt{arith} and \texttt{math} operations not supported by DCIR, thereby preventing unsupported operation errors from inundating our results and potentially masking significant findings.
The specific \texttt{arith} and \texttt{math} operations are not listed here for the sake of brevity.

\begin{figure}[t]
    \centering
    \includegraphics[width=\linewidth]{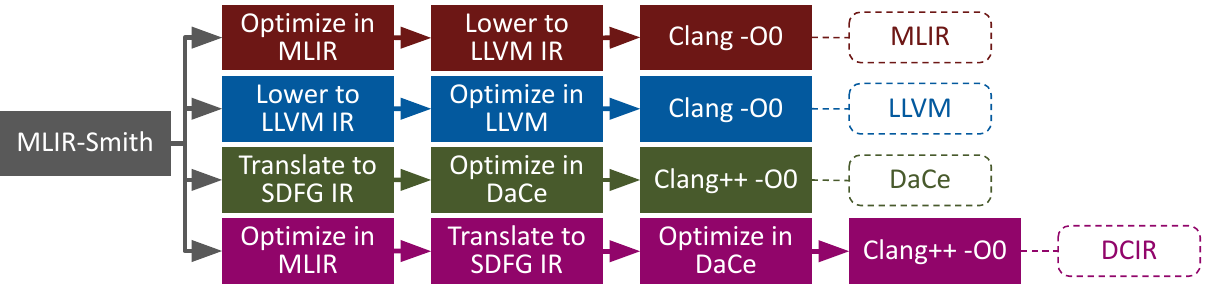}
    \vspace{-0.75em}
    \caption{Overview of the Pipelines used in the Evaluation.}
    \vspace{-1em}
    \label{fig:evaluation_pipelines}
\end{figure}

The generated MLIR code is processed through four distinct pipelines, as depicted in Fig.~\ref{fig:evaluation_pipelines}. We monitor for compilation errors and runtime discrepancies, including timeouts, print outputs, and exit codes. To isolate the optimization capabilities of each framework, we strategically disable optimizations in certain stages of the pipelines (e.g. with the \texttt{-O0} flag). The specifics of these pipelines are detailed in the following paragraphs.

\macsection{MLIR Pipeline}
In the MLIR pipeline, we utilize \texttt{mlir-opt}\footnote{\label{llvm-commit-mlir}Commit: 6f29d1adf29820daae9ea7a01ae2588b67735b9e} to execute dead code elimination (DCE), Common Subexpression Elimination (CSE), Loop Invariant Code Motion (LICM), function inlining and canonicalization. Subsequently, we lower to the LLVM dialect and generate LLVM IR using \texttt{mlir-translate}\footnoteref{llvm-commit-mlir}. The code is then compiled with \texttt{clang}\footnoteref{llvm-commit-mlir} using the flags \texttt{-O0 -fPIC -march=native}.

\macsection{LLVM Pipeline}
In the LLVM pipeline, we bypass MLIR optimizations, directly lowering to the LLVM dialect using \texttt{mlir-opt}\footnote{\label{llvm-commit}Commit: 6f29d1adf29820daae9ea7a01ae2588b67735b9e} and generating LLVM IR with \texttt{mlir-translate}\footnoteref{llvm-commit}. The code is then optimized with \texttt{opt}\footnoteref{llvm-commit} using the \texttt{-O3} flag and compiled with \texttt{clang}\footnoteref{llvm-commit} using the flags \texttt{-O0 -fPIC -march=native}.

\macsection{DaCe Pipeline}
In the DaCe pipeline, we convert the MLIR code to the SDFG dialect using \texttt{sdfg-opt}\footnote{\label{dcir-commit}Commit: 8734d2c10ecb9078a81ff3ae0b64a774b098aca5} and generate the SDFG IR using \texttt{sdfg-translate}\footnoteref{dcir-commit}. The code is then optimized using DaCe's auto-optimizer\footnoteref{dace-commit} and compiled with Clang++-11 (Ver. 11.1.0) using the \texttt{-O0} flag. We opted for Clang++-11 due to compatibility issues encountered with newer versions in the context of DaCe. It's important to note that Clang is solely used for compilation and not for optimization, ensuring that the results remain unaffected. This approach allows us to directly evaluate the performance and effectiveness of the DaCe framework in isolation.

\macsection{DCIR Pipeline}
In the DCIR pipeline, we initially perform the same optimizations as in the MLIR pipeline using \texttt{mlir-opt}\footnoteref{llvm-commit}, then convert it to the SDFG dialect using \texttt{sdfg-opt}\footnoteref{dcir-commit} and generate the SDFG IR using \texttt{sdfg-translate}\footnoteref{dcir-commit}. We then optimize using DaCe's auto-optimizer\footnote{\label{dace-commit}Commit: 4fcb0b5ee90384829ab7a76bde0a07a7a8cbcb4b} and compile the optimized SDFG with Clang++-11 (Ver. 11.1.0) using the \texttt{-O0} flag. Our choice of Clang++-11 aligns with the rationale provided in the DaCe pipeline section.

\begin{table}[t]
  \centering
  \footnotesize
  \caption{Configuration parameters for MLIR-Smith in the comparative evaluation of MLIR, LLVM, DaCe, and DCIR compiler pipelines.}
  \vspace{-1em}
    \begin{tabularx}{\linewidth}{@{} p{8em} p{3em} *1{>{\raggedright\arraybackslash}X} @{}}
        \toprule
            \textbf{Parameter} & \textbf{Value} & \textbf{Description} \\
        \midrule
            \texttt{regionDepthLimit} & \texttt{4} & The maximal depth of nested regions. \\
            
            \addlinespace[0.5em]
            \texttt{blockLength} & \texttt{50} & The maximal amount of operations inside one block. \\
            
            \addlinespace[0.5em]
            \texttt{defaultProb} & \texttt{1} & The default probability value for an operation. \\
            
            \addlinespace[0.5em]
            \texttt{scf.if} & \texttt{10} & Probability value for the \texttt{scf.if} operation. \\

            \addlinespace[0.5em]
            \texttt{scf.for} & \texttt{10} & Probability value for the \texttt{scf.for} operation. \\
            
            \addlinespace[0.5em]
            \texttt{scf.while} & \texttt{10} & Probability value for the \texttt{scf.while} operation. \\
        \bottomrule
    \end{tabularx}
  \label{table:evaluation_configparams}
  \vspace{-1em}
\end{table}

\subsection{Uncovered Bugs and Missed Optimizations}
Throughout our evaluation process, as delineated in \S\ref{sec:experimental_setup}, we encountered a range of bugs across the different compiler pipelines. In this subsection, we have handpicked and present a selection of the most notable bugs that surfaced. We delve into a comprehensive exploration of their characteristics, the specific circumstances that led to their detection, and the impact they exert on their respective compiler pipelines.

\begin{figure}[t]
    \centering
    \begin{subfigure}{.8\linewidth}
\begin{lstlisting}[style=tinyMLIR]
module {
  func.func @main() -> i32 {
    %alloc = memref.alloc() : memref<0xi32>
    %c0 = arith.constant 0 : index
    %0 = memref.load %alloc[%c0] : memref<0xi32>
    %c0_i32 = arith.constant 0 : i32
    %1 = arith.maxsi %0, %c0_i32 : i32
    return %c0_i32 : i32
  }
}
\end{lstlisting}
        \vspace{-0.5em}
        \caption{Reduced MLIR program revealing DCE bug in DaCe.}
        \label{fig:evaluation_dace_dce_code}
    \end{subfigure}
    \begin{subfigure}{\linewidth}
        \centering
        \includegraphics[width=.8\textwidth]{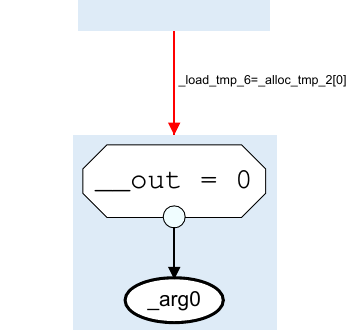}
        \caption{SDFG after optimization by DaCe.}
        \label{fig:evaluation_dace_dce_opt}
    \end{subfigure}
    \vspace{-1em}
    \caption{Illustration of the DCE bug in DaCe.}
    \vspace{-1em}
    \label{fig:evaluation_dace_dce}
\end{figure}

\macsection{DaCe Bug}
Our evaluation process led to the discovery of a dead code elimination (DCE) bug in DaCe, as demonstrated by the reduced MLIR program shown in Fig.~\ref{fig:evaluation_dace_dce_code}. Despite successfully eliminating most operations, DaCe fails to identify that \texttt{\_load\_tmp\_6} remains unused in the bottom state, as illustrated in Fig.~\ref{fig:evaluation_dace_dce_opt}. This oversight hinders the elimination of the edge assignment, which subsequently prevents the removal of the array allocation. This bug persists irrespective of multiple invocations and different optimizer configurations, such as reordering the simplification and auto-optimization passes. 
Interestingly, the removal of the \texttt{arith.maxsi} operation results in correct compilation and execution. Our investigations revealed that the SDFG generated by \texttt{sdfg-translate} is identical except for the \texttt{arith.maxsi} operation, suggesting that the bug resides in the DaCe optimizer. Upon reporting this bug to the DaCe development team, they were able to reproduce the issue and subsequently acknowledged its validity.

\macsection{DCIR Bug}
In the course of our evaluation, we detected translation bugs in DCIR, specifically related to the \texttt{arith.extsi} operation for boolean values. Fig.~\ref{fig:evaluation_dace_dce_code} displays the reduced MLIR program that led to the unearthing of this translation bug in DCIR. As shown in Fig.~\ref{fig:evaluation_dace_dce_opt}, DCIR translates the \texttt{arith.extsi} operation as a mere copy from the input to the output. While this translation is accurate for an input boolean of \emph{false}, a \emph{true} input value undergoing a signed integer extension should result in an output integer with each bit set to $1$. However, DCIR incorrectly treats all signed integer extensions as unsigned, leading to incorrect results. Further investigations revealed that the translation of \texttt{arith.extsi} is generally flawed for negative integers in DCIR.
We attempted to further reduce the input MLIR program, but simpler versions led to \texttt{sdfg-opt} performing constant folding and propagation before translating with \texttt{sdfg-translate}, which concealed the bug. This bug was reported and subsequently confirmed by the DCIR development team.

\begin{figure}[t]
    \centering
    \begin{subfigure}{.8\linewidth}
\begin{lstlisting}[style=tinyMLIR]
module {
  func.func @main() -> i32 {
    %c0_i8 = arith.constant 0 : i8
    %0 = arith.extsi %c0_i8 : i8 to i64
    %1 = arith.muli %0, %0 : i64
    %2 = arith.cmpi ule, %0, %1 : i64
    %3 = arith.extsi %2 : i1 to i32
    %4 = arith.addi %3, %3 : i32
    return %4 : i32
  }
}
\end{lstlisting}
        \vspace{-0.5em}
        \caption{Reduced MLIR program revealing the bug in DCIR.}
        \label{fig:evaluation_dcir_extsi_code}
    \end{subfigure}
    \begin{subfigure}{\linewidth}
        \centering
        \includegraphics[width=.5\textwidth]{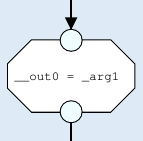}
        \caption{Translated \texttt{arith.extsi} tasklet in the resulting SDFG.}
        \label{fig:evaluation_dcir_extsi_opt}
    \end{subfigure}
    \vspace{-1em}
    \caption{Illustration of the translation bug in DCIR.}
    \vspace{-1em}
    \label{fig:evaluation_dcir_extsi}
\end{figure}

\macsection{MLIR Missed Optimization}
Our evaluation process, as detailed in \S\ref{sec:experimental_setup}, led us to the discovery of a missed optimization within the MLIR pipeline. This overlooked optimization becomes apparent when processing the MLIR program illustrated in Figure~\ref{fig:evaluation_mlir_missed_code} through the MLIR pipeline. Executing the generated program results in a segmentation fault, which we attribute to a large, non-eliminated memory allocation that remains unutilized in the input code. The likely failure of this memory allocation subsequently triggers a null pointer dereference during the read operation. This assertion is further supported by an inspection of the generated assembly (refer to Figure~\ref{fig:evaluation_mlir_missed_asm}), where it's evident that all memory operations persist.

Remarkably, the MLIR pipeline demonstrates an ability to eliminate the memory allocation if either the load or store operations are removed from the input code. Moreover, using the read value in other operations or storing a value from another operation also leads to the memory allocation's elimination. This finding suggests that while the individual operations do not present an issue and are subject to elimination, the dependence between load and store operations seems to hinder dead code elimination, resulting in the aforementioned segmentation fault. Notably, all other pipelines were able to identify the unused memory and proceeded to eliminate the memory allocation.


In conclusion, the evaluation of MLIR-Smith underscores its utility in identifying bugs and missed optimizations across multiple compiler pipelines. These findings reinforce the necessity of comprehensive testing in enhancing the reliability and capability of such systems.
\section{Related Work}
In this section, we delve into an exploration of alternative methodologies that exhibit similarities in objectives or techniques to our own approach.

\begin{figure}[t]
    \centering
    \begin{subfigure}{.8\linewidth}
\begin{lstlisting}[style=tinyMLIR]
module {
  func.func @main() -> i32 {
    %c0 = arith.constant 0 : index
    %alloc = memref.alloc() : memref<100000x100000xi64>
    %0 = memref.load %alloc[%c0,%c0] : memref<100000x100000xi64>
    memref.store %0, %alloc[%c0,%c0] : memref<100000x100000xi64>
    memref.dealloc %alloc : memref<100000x100000xi64>
    %c2_i32 = arith.constant 2 : i32
    return %c2_i32 : i32
  }
}
\end{lstlisting}
        \vspace{-0.5em}
        \caption{Reduced MLIR program revealing the missed optimization in MLIR.}
        \label{fig:evaluation_mlir_missed_code}
    \end{subfigure}
    \vskip\baselineskip
    \begin{subfigure}{.8\linewidth}
\begin{lstlisting}[style=tinyASM]
main:
  pushq	%rax
  xorl	%eax, %eax
  movl	%eax, %edi
  movabsq	$80000000000, %rax
  addq	%rax, %rdi
  callq	malloc@PLT
  movq	%rax, %rdi
  movq	(%rdi), %rax
  movq	%rax, (%rdi)
  callq	free@PLT
  movl	$2, %eax
  popq	%rcx
  retq
\end{lstlisting}
        \vspace{-0.5em}
        \caption{Generated assembly code of the MLIR pipeline.}
        \label{fig:evaluation_mlir_missed_asm}
    \end{subfigure}
    \vspace{-1em}
    \caption{Illustration of the missed optimization in MLIR.}
    \label{fig:evaluation_mlir_missed}
\end{figure}

\macsection{Historical Overview}
The history of compiler testing using random techniques spans over half a century. Boujarwah and Saleh~\cite{boujarwah1997compiler} conducted a comprehensive exploration of this field in 1997. Richard L. Sauder~\cite{10.1145/1460833.1460869} pioneered this area in 1962 when he applied random variables in program data sections to evaluate the performance of COBOL compilers. His work was centered around the development of a "Test Data Generator" that worked in conjunction with the COBOL compiler. This system was designed to not only generate data that conformed to the descriptions given in the Data Division of a COBOL program but also to establish necessary data relationships to test the logic of the COBOL program.

In 1970, building on Sauder's work, K.V. Hanford~\cite{10.1147/sj.94.0242} developed a system for the automatic generation of test cases. This system was designed to test the reliability of programming products in accepting new test cases without any errors. The input to the system was a syntax definition in a formal notation, which was a challenging task but provided valuable insights into the structure of the language and highlighted any obscurities or ambiguities in the existing documentation. Hanford's system had definitions for ECMA Algol, FORTRAN IV, and a major subset of PL/I.
%

In 1972, Purdom~\cite{10.1007/BF01932308} introduced a syntax-guided approach for generating test sentences for parsers. In his publication Purdom outlined a technique that employed an efficient algorithm capable of creating a concise collection of brief sentences derived from a context-free grammar. This approach guaranteed the utilization of every production rule within the grammar at least once. These sentences were beneficial for testing parsing programs and debugging grammars. The sentences were also used to test some automatically generated simpleLR(1) parsers, demonstrating the effectiveness of this approach for testing LR(1) parsers.

\macsection{Random Generation and Automated Testing}
Barany~\cite{barany2017livenessdriven} introduced a novel approach to random program generation, which he called "liveness-driven". This method constructs the random program from the bottom-up, guided by a simultaneous structural data-flow analysis to ensure that the generator never produces dead code. Barany's approach was implemented as a plugin for the Frama-C framework. In comparison to Csmith, the standard random C program generator, Barany's tool was found to generate programs that compile to more machine code with a more complex instruction mix. Contrary to our approach, this approach was particularly effective in avoiding the generation of dead code. 
%

Pałka et al.~\cite{10.1145/1982595.1982615} introduced a method for random testing compilers, using randomly generated programs as inputs, and comparing their behavior with and without optimization. They focused on generating typed functions on lists, which were compiled using the Glasgow Haskell compiler, a mature production quality Haskell compiler. After around 20,000 tests, they triggered an optimizer failure, and automatically simplified it to a program with just a few constructs. Their approach was unique in that it took into account syntax, scope rules, and type checking during random generation, while attaining a good distribution of test data.


%


Finally, Yang et al.~\cite{Csmith} introduced Csmith, a tool capable of generating programs that encompass a large portion of C while steering clear of undefined and unspecified behaviors. Their approach combined static analysis and dynamic checks to limit undefined or unspecified behaviors. Utilizing Csmith, they identified over 325 faults in GCC~\cite{gcc} and LLVM~\cite{llvm}. 
They also provided a comprehensive set of qualitative and quantitative results about the bugs they discovered in open-source C compilers.

Our work distinguishes itself from the previously discussed approaches by adopting a fundamentally different methodology. Rather than focusing on a specific language with a pre-defined set of operations, we have designed an extensible framework for code generation. This framework is primarily aimed at facilitating fuzz and differential testing for developers working on new languages, with a particular emphasis on dialects of MLIR.

\macsection{Incorporation of Advanced Techniques}
Our framework is designed to easily incorporate advanced techniques. For instance, the optimization markers proposed by Theodoridis et al.~\cite{10.1145/3503222.3507764} can be seamlessly integrated into our system. The contribution of Theodoridis et al. is of particular significance due to their innovative application of optimization markers, which facilitate the identification of overlooked optimizations through the perspective of dead code elimination. Their methodology led to the unearthing of 84 bugs in GCC~\cite{gcc} and LLVM~\cite{llvm}, 62 of which have already been confirmed or rectified.

In summary, our work marks a substantial shift from conventional methods, providing a more adaptable and dynamic solution for compiler testing. By capitalizing on the extensibility of our framework, we strive to render the process of compiler testing and the exploration of new techniques more accessible and efficient for developers.

\section{Conclusion}
This paper introduces MLIR-Smith, a novel tool for generating random programs in the MLIR framework. It demonstrates the tool's effectiveness through a series of evaluations, showcasing its potential to enhance the field of compiler research and testing. The tool is designed to be highly extensible and configurable, allowing developers to easily integrate new operations from new dialects and customizing crucial aspects of the tool to meet their specific requirements.
The scope of this work can be broadened in the future by incorporating additional core MLIR dialects into MLIR-Smith, thereby expanding its scope. Additionally, operations should have the capability to add their distinct configurations to MLIR-Smith.
Finally, introducing support for composite types and unbounded sets of types would provide a significant advantage.

%
%
%
%
%






\appendix
\section{Artifact Appendix}

\subsection{Abstract}

The artifacts associated with this paper include the MLIR-Smith tool and the experimental setup, as detailed in \S\ref{sec:experimental_setup}. This is supplemented with scripts designed to streamline the experimental workflow, as well as the configuration file referenced in \S\ref{sec:experimental_setup}. The experiment is designed to operate on standard hardware, and for convenience, a Docker container has been provided to simplify the setup process. It's important to note that all software utilized is publicly accessible and has been included within the provided container.

\subsection{Artifact Check-list (Meta-information)}


{\small
\begin{itemize}
  \item {\bf Compilation:} Clang (Ver. 17.0.0), Clang-11 (Ver. 11.1.0), DaCe (git revision \texttt{4fcb0b5})\footnote{\url{https://github.com/Berke-Ates/dace/tree/4fcb0b5}}
  \item {\bf Transformations:} MLIR Optimizer and Translator (git revision \texttt{f4dfabd})\footnote{\label{arti_llvm}\url{https://github.com/Berke-Ates/llvm-project/tree/f4dfabd}}, LLVM Optimizer and Compiler (git revision \texttt{00a1258})\footnoteref{arti_llvm}, DCIR Optimizer and Translator (git revision \texttt{8734d2c})\footnote{\url{https://github.com/spcl/mlir-dace/tree/8734d2c}}
  \item {\bf Binary:} MLIR-Smith Program Generator (git revision \texttt{f4dfabd})\footnoteref{arti_llvm}
  \item {\bf Run-time environment:} Tested on Ubuntu 20.04 and 22.04. Depending on Docker installation, root access may be required.
  \item {\bf Metrics:} Compilation errors, runtime differences, missed DCE optimizations
  \item {\bf Output:} Generated code and all metrics sorted in subfolders
  \item {\bf Experiments:} Several scripts to execute the experiments are included
  \item {\bf How much disk space required (approximately)?:}\\676~MiB to pull Docker container (2.1~GiB uncompressed)
  \item {\bf How much time is needed to complete experiments (approximately)?:} 5 minutes
  \item {\bf Publicly available?:} Yes
  \item {\bf Code licenses (if publicly available)?:} BSD 3-Clause License
  \item {\bf Workflow framework used?:} No
  \item {\bf Archived (provide DOI)?:} No
\end{itemize}
}

\subsection{Description}

\subsubsection{How to Access}
All necessary files, along with instructions for their execution, can be found on GitHub (\url{https://github.com/Berke-Ates/AST}).



\subsubsection{Software dependencies}
In order to build and run the Docker container, it's necessary to have Docker installed on your system, along with an active instance of the Docker daemon.


\subsection{Installation}

Instructions for both manual and dockerized installations are detailed in the \texttt{README.md} file located in the repository. The dockerized installation is recommended for ease of use and can be initiated with the following command: 

\texttt{docker pull berkeates/mlir-smith:latest}

\subsection{Experiment workflow}

In our experiments, we start by using \texttt{mlir-smith} to generate a random MLIR program. This program is then processed through the pipelines as outlined in \S\ref{sec:experimental_setup}. We monitor for any compiler errors during this process. If an error is detected, the execution is halted and the MLIR input file, along with all associated logs, is moved to a separate folder named \texttt{comp\_err}.
If all compilers successfully process the program, we then execute the generated binaries and compare both the print outputs and the exit codes. If there are discrepancies, the execution is again halted and all relevant files are moved to a folder named \texttt{exe\_diff}.
In cases where the observable behavior is consistent across all compiler pipelines, we proceed to examine the external function calls in the assembly to identify any missed opportunities for dead code optimization. If we find that at least one compiler has eliminated more external function calls, the files are moved to a folder named \texttt{flag\_diff}.
If all these checks pass without any issues, indicating no detectable differences, the files are placed in the \texttt{normal} folder.

All the necessary compilers, translators, and code generators are included and built within the Docker container. This container can be initiated with the command:

\texttt{docker run -it berkeates/mlir-smith}.

\subsection{Evaluation and Expected Result}


The \texttt{diff\_test.sh} script, located within the \texttt{scripts} folder, is designed to invoke MLIR-Smith to randomly generate MLIR programs and process them through the four pipelines detailed in \S\ref{sec:experimental_setup}. This script requires a path to MLIR-Smith, an output directory, and an optional config file.
An example of how to execute this script might look like the following: \texttt{./scripts/diff\_test.sh ./bin/mlir-smith out diff\_test.config}. The results of this process can be reviewed in the newly created \texttt{./out} folder.

\subsection{Experiment Customization}

The configuration can be modified by either editing the file named \texttt{diff\_test.config} in the home directory or by supplying a different config file to the \texttt{diff\_test.sh} script. If you wish to examine a specific input file, you can perform a single differential test by executing the \texttt{diff\_test\_file.sh} script, which is located in the \texttt{scripts} folder.







\bibliographystyle{ACM-Reference-Format}
\bibliography{refs}

\end{document}